\documentclass[a4paper,12pt]{article}

\usepackage[usenames]{color}
\usepackage{graphicx}
\usepackage{amsmath}
\usepackage{amssymb}
\usepackage[normalem]{ulem}

\newcommand {\cD}{{\cal D}}

\newcommand {\cN}{{\cal N}}

\makeatletter
\@addtoreset{equation}{section}
\makeatother

\newcommand{\be}{\begin{equation}}
\newcommand{\ee}{\end{equation}}
\newcommand{\bea}{\begin{eqnarray}}
\newcommand{\eea}{\end{eqnarray}}

\def\double #1{#1{\hbox{\kern-2pt $#1$}}}

\renewcommand\sout{\bgroup \color{red} \ULdepth=-.5ex \ULset}

\begin{document}
\begin{titlepage}
\null

\vskip 1.8cm
\begin{center}
 
  {\Large \bf Non-Abelian Chern-Simons Actions \\
 \vskip 0.2cm 
in Three-dimensional Projective Superspaces
}

\vskip 1.8cm
\normalsize
\renewcommand\thefootnote{\alph{footnote}}

  {\bf Masato Arai$^{\dagger}$\footnote{masato.arai(at)fukushima-nct.ac.jp} 
and Shin Sasaki$^\sharp$\footnote{shin-s(at)kitasato-u.ac.jp}}

\vskip 0.5cm

  { \it 
  $^\dagger$Fukushima National College of Technology \\
   Iwaki, Fukushima 970-8034, Japan \\
  \vskip 0.2cm
  $^\dagger$
  Institute of Experimental and Applied Physics, \\
  Czech Technical University in Prague, \\
  Horsk\' a 3a/22, 128 00, Prague 2, Czech Republic \\
  \vskip 0.2cm
  $^\sharp$Department of Physics \\
  Kitasato University \\
  Sagamihara, 252-0373, Japan
  }

\vskip 2cm

\begin{abstract}
We construct an action for the superconformal Chern-Simons theory 
with non-Abelian gauge groups in three-dimensional $\mathcal{N} = 3$ projective superspace.
We propose a Lagrangian 
given by the product of the function of the tropical multiplet,
that represents the $\mathcal{N} = 3$ vector multiplet, 
and the $\mathcal{O}(-1,1)$ multiplet.
We show how the tropical multiplet is embedded into the $\mathcal{O}(-1,1)$ 
multiplet by comparing our Lagrangian with the 
Chern-Simons Lagrangian in the $\mathcal{N} = 2$ superspace.
We also discuss $\mathcal{N} = 4$ generalization of the action.
\end{abstract}

\end{center}
\end{titlepage}
\newpage
\setcounter{footnote}{0}

\section{Introduction}
Off-shell superfield formalisms of supersymmetric gauge theories with eight supercharges 
have been studied in the past.
There are two known off-shell formalisms with eight supercharges.
One is the harmonic superspace formalism \cite{GaIvOgSo, GaIvKaOgSo, HSS-B}, 
the other is the projective superspace formalism \cite{KaLiRo, LiRo1}.

Recently, the projective superspace formalism that keeps manifest $\mathcal{N} = 3$ 
and $\mathcal{N} = 4$ superconformal symmetries in three dimensions has been 
developed \cite{KuPaTaUn}. 
This results open up a window to construct a supersymmetric Chern-Simons actions
in terms of projective superfields. The harmonic superspace provides an explicit form of 
the $\cN=3$ non-Abelian Chern-Simons actions \cite{ZuKh, Zu} while it
had not been constructed in three-dimensional projective superspace formalisms.
In \cite{ArSa1}, we have constructed the $\mathcal{N} = 3$ and $\mathcal{N} = 4$
superconformal Chern-Simons actions with Abelian gauge group in the projective superspace.
We have shown that the $\cN=3$ Abelian Chern-Simons Lagrangian is written in the 
product of the tropical multiplet representing an Abelian vector multiplet $\mathcal{V}^{[0]}$
with weight 0 and an $\mathcal{O} (-1,1)$ multiplet $G^{[2]}$ with
weight 2 corresponding to the field strength of the vector multiplet\footnote{
A brief introduction of the projective superfield formalism is found in
Appendix B.
}. 
The $\mathcal{N} = 3$ vector multiplet 
consists of a (anti)chiral superfield $\Phi$
($\bar{\Phi}$) and a vector superfield $V$ in the $\mathcal{N} = 2$ standard superfield formalism. We have found the explicit relations among the $\mathcal{N} = 2$ superfields $(V, \Phi, \bar{\Phi})$ and 
the $\mathcal{N} = 3$ projective superfield $\mathcal{V}^{[0]}$
and determined the embedding of the vector multiplet into the tropical multiplet.
This is the key point to obtain the $\cN=3$ superconformal Chern-Simons actions in the
projective superspace formalism since the action is constructed so that it reproduces
the known Chern-Simons action in terms of $\cN=2$ superfields. 
However, generalizing this result to that with non-Abelian gauge groups is not straightforward.
For non-Abelian gauge groups, the relation between the tropical multiplet 
$\mathcal{V}^{[0]}$ and the $\cN=3$ vector multiplets in terms of $\cN=2$ 
superfields have not been studied in detail.
This is because the relation between them becomes non-linear and complicated.
Formal treatments of non-Abelian vector multiplets in four dimensions
have been discussed for example in \cite{LiRo2, Go, GoRo}.
In \cite{ArSa2}, we have studied three-dimensional 
$\mathcal{N} = 3$, $\mathcal{N} = 4$ and four-dimensional $\mathcal{N} =
2$ charged hypermultiplets that couple with the non-Abelian vector multiplet and
found the relations among the non-Abelian vector multiplet ($V, \Phi, \bar{\Phi}$) 
and the $\mathcal{N} = 3$ tropical multiplet $\mathcal{V}^{[0]}$.
We have explicitly written down the $\mathcal{N} = 2$ superfields ($V, \Phi,
\bar{\Phi}$) as functions of the components in $\mathcal{V}^{[0]}$.
The same analysis has been done for the $\mathcal{N} = 4$ and
the four-dimensional $\mathcal{N} = 2$ cases.
We also constructed the actions of hypermultiplets coupled with the non-Abelian 
vector multiplet in the three and four-dimensional projective superspaces.

The purpose of this paper is to construct the non-Abelian Chern-Simons action in the
three dimensional $\mathcal{N} = 3$ projective superspace.
Although the action has been discussed in the harmonic superspace formalism, our study provides a complimentary analysis of
the non-Abelian Chern-Simons theory in the projective superspace formalism.
The Lagrangian is written in the form of the product of ``a gauge field $f(\mathcal{V}^{[0]})$''
and ``a gauge field strength $G^{[2]}$''.
In order to have the consistent Abelian limit, 
the gauge field strength $G^{[2]}$ should belong to the $\mathcal{O}
(-1,1)$ multiplet. We will show that $G^{[2]}$ actually satisfies the
condition of $\mathcal{O} (-1,1)$ multiplet.
We also find the explicit form of the function $f$ which is consistent
with the Abelian limit.
Finally we will show that the action proposed in this letter completely
reproduces the known action for the non-Abelian Chern-Simons theory in the $\mathcal{N} = 2$ standard 
superfield formalism.

The organization of this letter is as follows. 
In the next section, we give a brief review about the embedding of the
$\mathcal{N} = 3$ non-Abelian vector multiplet into the tropical multiplet in the
projective superspace formalism.
We work in the formalism that the superconformal symmetry is manifest
\cite{KuPaTaUn}.
In Section 3, we propose the non-Abelian Chern-Simons action in the
$\mathcal{N} = 3$ projective superspace.
We show that the action in the projective superspace 
reproduces that in the $\mathcal{N} = 2$ superspace.
We also discuss the $\mathcal{N} = 4$ generalization of the action.
Section 4 is conclusion and discussions.
Notations and conventions of the three-dimensional $\mathcal{N} = 2, 3, 4$ 
ordinary superspaces are given in Appendix A. 
A brief summary of the three-dimensional $\mathcal{N} = 3$ and the $\mathcal{N}
= 4$ projective superspace formalisms are found in Appendix B.

\section{Non-Abelian vector multiplet in projective superspace}
In this section, we give a brief overview 
about non-Abelian vector multiplets in the projective superspace formalism.
A summary of the three-dimensional $\mathcal{N} = 3$ and $\mathcal{N} = 4$ projective
superspace formalism is found in Appendix B. 
For more detail of the formalism, see \cite{KuPaTaUn, ArSa1, ArSa2}.

The non-Abelian vector multiplet is 
represented as the real tropical multiplet $\mathcal{V}^{[0]}$ with weight 0.
In the Lindstr\"om-Ro\v{c}ek gauge, $\mathcal{V}^{[0]}$ is expanded by the
projective coordinate $\zeta$ parameterizing $\mathbb{C}P^1$ as \cite{LiRo2} 
\begin{align}
& \mathcal{V}^{[0]}(z,\zeta) = \frac{1}{\zeta} V_{-1}(z) + V_0(z) + \zeta V_1(z), \notag \\
& \bar{V}_0(z) = V_0(z), \qquad \bar{V}_1(z) = - V_{-1}(z),
\end{align}
where $z$ is the $\cN=3$ standard superspace coordinate defined in Appendix A. 
The component superfields $V_{-1}, V_0, V_1$ are adjoint representation
of a non-Abelian gauge group.
The tropical multiplet satisfies the projective superfield constraint
$D^{[2]}_{\alpha} \mathcal{V}^{[0]} = 0$. 
This is equivalent to the following constraints on the component superfields:
\begin{eqnarray}
\begin{aligned}
 & \mathbb{D}_{\alpha} V_1 = 0, \\
 & \mathbb{D}_{\alpha} V_0 - 2 D^{12}_{\alpha} V_1 = 0, \\
 & \mathbb{D}_{\alpha} V_{-1} - 2 D^{12}_{\alpha} V_0 -
 \bar{\mathbb{D}}_{\alpha} V_1 = 0, \\
 & \bar{\mathbb{D}}_{\alpha} V_0 + 2 D^{12}_{\alpha} V_{-1} = 0, \\
 & \bar{\mathbb{D}}_{\alpha} V_{-1} = 0.
\end{aligned}
\label{eq:tropical_constraint}
\end{eqnarray}
Following the analysis in 
four dimensions \cite{LiRo1}, we decompose the
tropical multiplet as follows:
\begin{align}
e^{\mathcal{V}^{[0]}} = e^{\hat{V}_{-}} e^{\hat{V}_0} e^{\hat{V}_{+}},
\label{eq:decomp}
\end{align}
where $\hat{V}_{+}$ ($\hat{V}_{-}$) contains only positive (negative)
powers of $\zeta$ and $\hat{V}_0$ contains terms with $\zeta^0$.
In general, they are expanded by $\zeta$ as 
\begin{align}
\hat{V}_{+} = \sum_{n=1}^{\infty} \zeta^n \hat{V}_n, \qquad 
\hat{V}_{-} = \sum_{n=1}^{\infty} \zeta^{-n} \hat{V}_{-n},
\end{align}
where $\hat{V}_{\pm n}$ and 
$\hat{V}_0$ are functions of $V_{\pm 1}, V_0$. 
Finding the closed expressions of the functions $\hat{V}_{\pm n},
\hat{V}_0$ in terms of the components in the tropical multiplet $\mathcal{V}^{[0]}$
is difficult in the projective superspace formalism.
However they are obtained perturbatively.
Up to $\mathcal{O} (V^4)$, $\hat{V}_{\pm n}, \hat{V}_0$ are found to be, 
\begin{align}
\hat{V}_{-2} &= 
\frac{1}{12} 
\left[
V_{-1}, [V_0, V_{-1}]
\right] + \mathcal{O} (V^4), \notag \\
\hat{V}_{-1} &= V_{-1} + \frac{1}{2} [V_0, V_{-1}] 
+ \frac{1}{6} 
\left[
V_{-1}, [V_{-1}, V_1]
\right]
+ \frac{1}{6}
\left[
V_0, [V_0, V_{-1}]
\right]
+ \mathcal{O} (V^4), \notag \\
\hat{V}_0 &= V_0 
+ \frac{1}{2} [V_1, V_{-1}] 
- \frac{1}{12} 
\left[
V_1, [V_{-1}, V_0] 
\right] 
- \frac{1}{12} 
\left[
V_{-1}, [V_1, V_0]
\right]
+ \mathcal{O} (V^4) , \notag \\
 \hat{V}_1 &= V_1 + \frac{1}{2} [V_1, V_0]
 + \frac{1}{6} 
 \left[
 V_1, [V_1, V_{-1}]
 \right]
 + \frac{1}{6}
 \left[
 V_0, [V_0, V_1]
 \right]
 + \mathcal{O} (V^4), \notag  \\
 \hat{V}_{2} &= 
 \frac{1}{12} 
 \left[
 V_1, [V_0, V_1]
 \right]
 + 
\mathcal{O} (V^4).
\label{eq:potential}
\end{align}
We note that in the Abelian limit, these components become $\hat{V}_{\pm 1} = V_{\pm
1}$, $\hat{V}_0 = V_0$ and $\hat{V}_{\pm n} = 0 \ (n > 2)$.
As we will see in below, even for 
the non-Abelian gauge group, the components $\hat{V}_{n} \ (n > 2)$,
$\hat{V}_{-n} \ (n > 3)$ 
do not appear in the Lagrangian 
and they are not relevant to our discussion.

In order to express the $\cN=3$ vector multiplet $(V, \Phi, \bar{\Phi})$ 
in terms of the component superfields $V_{-1}, V_0, V_1$ in the tropical
multiplet $\mathcal{V}^{[0]}$, we consider the action for
hypermultiplets 
that couples to tropical multiplets. 
The hypermultiplets charged under the non-Abelian gauge group 
are embedded in the (ant)arctic multiplets $\Upsilon^{[1]}$($\bar{\Upsilon}^{[1]}$) with weight 1, 
which are expressed as
\begin{eqnarray}
\Upsilon^{[1]}(z,\zeta)= \sum^{\infty}_{n=0} \zeta^n \Upsilon_n (z),\quad 
\bar{\Upsilon}^{[1]} (z,\zeta)=\sum^{\infty}_{n=0} \left(-{1 \over \zeta}
\right)^n \bar{\Upsilon}_n (z). \label{expand}
\end{eqnarray}
They also satisfy the projective superfield constraints 
$D^{[2]}_{\alpha} \Upsilon^{[1]} = D^{[2]}_{\alpha} \bar{\Upsilon}^{[1]}=0$.

The Lagrangian for the hypermultiplet in the fundamental representation of the 
non-Abelian gauge group is originally written in the $\cN=3$ superspace.
After choosing a frame where the isospinor $u$ is fixed, the Lagrangian
becomes (see (\ref{N2action})):
\begin{align}
\mathcal{L} = \frac{1}{2\pi i} \oint_\gamma \! \frac{d \zeta}{\zeta} d^4 \theta
 \ \bar{\Upsilon}^{[1]} e^{\mathcal{V}^{[0]}} \Upsilon^{[1]},
\label{eq:hyper}
\end{align}
where the contour $\gamma$ is chosen such that it does not through the north pole
of $\mathbb{C}P^1$.
We can reduce this Lagrangian to that in terms of $\cN=2$ superfields.
Below we briefly explain how to reduce it and how to find the relation between 
the component superfields $V_{-1}, V_0, V_1$ in ${\cal V}^{[0]}$ and the $\cN=2$ superfields $(V, \Phi, \bar{\Phi})$. First we define the following new fields
\begin{eqnarray}
\bar{\tilde{\Upsilon}}^{[1]} \equiv \bar{\Upsilon}^{[1]} e^{\hat{V}_{-}},
 \qquad 
\tilde{\Upsilon}^{[1]} \equiv e^{\hat{V}_0} e^{\hat{V}_{+}}
\Upsilon^{[1]},
\label{eq:N3_tilde}
\end{eqnarray}
which satisfy the gauge-covariantized projective superfield constraints:
\begin{eqnarray}
 \cD_\alpha^{[2]}\tilde{\Upsilon}=0 \label{const-p}.
\end{eqnarray}
Here $\cD_\alpha^{[2]}$ is the gauge covariant derivative defined to be 
\begin{align}
& \mathcal{D}^{[2]}_{\alpha} = - \bar{\mathcal{D}}_{\alpha} - 2 \zeta
 \mathcal{D}^{12}_{\alpha} + \zeta^2 \mathcal{D}_{\alpha}, \\
& \mathcal{D}_{\alpha} = \mathbb{D}_{\alpha} + \Gamma^{(-)}_{-2 \alpha},
 \quad 
\mathcal{D}^{12}_{\alpha} = D^{12}_{\alpha} - \frac{1}{2}
 \Gamma^{(-)}_{-1 \alpha}, 
\quad 
\bar{\mathcal{D}}_{\alpha} = \bar{\mathbb{D}}_{\alpha} - \Gamma^{(-)}_{0
 \alpha},
\end{align}
where the gauge connections are defined as \cite{ArSa2}
\begin{align}
\Gamma^{(-)}_{-2 \alpha} = 0, \quad 
\Gamma^{(-)}_{-1 \alpha} = \mathbb{D}_{\alpha} \hat{V}_{-1}, \quad 
\Gamma^{(-)}_{0 \alpha} = \mathbb{D}_{\alpha} \hat{V}_{-2} - 2
 D^{12}_{\alpha} \hat{V}_{-1} + \frac{1}{2} [\mathbb{D}_{\alpha}
 \hat{V}_{-1}, \hat{V}_{-1}].
\end{align}
The algebra that the gauge covariant derivatives $\mathcal{D}_{\alpha}$,
$\mathcal{D}^{12}_{\alpha}$, $\bar{\mathcal{D}}_{\alpha}$ satisfy was calculated
in \cite{ArSa2}.

With the use of  (\ref{eq:N3_tilde}), (\ref{eq:hyper}) is simplified to be 
\begin{align}
\mathcal{L} = \frac{1}{2\pi i} \oint_\gamma \! \frac{d \zeta}{\zeta} d^4 \theta
 \ \bar{\tilde{\Upsilon}}^{[1]} \tilde{\Upsilon}^{[1]}.
\label{eq:hyper2}
\end{align}
This is a convenient form to write down the Lagrangian in terms of the $\cN=2$ superfields since 
$\tilde{\Upsilon}^{[1]}$ and $\bar{\tilde{\Upsilon}}^{[1]}$ can be expanded 
as the same way in (\ref{expand}): 
One finds the expanded forms by replacing $\Upsilon_n$ with $\tilde{\Upsilon}_n$ and so on.
Substituting their expanded forms into (\ref{eq:hyper2}) and integrating over $\zeta$, 
we are left with $\tilde{\Upsilon}_0$ and $\tilde{\Upsilon}_1$ in the Lagrangian. 
The other fields $\tilde{\Upsilon}_n (n\ge 2)$ are integrated out since they are 
non-dynamical. Equation (\ref{const-p}) gives a constraint for  $\tilde{\Upsilon}_1$
and it is incorporated into the Lagrangian with the Lagrange multiplier $\cN=2$ 
superfield. Then integrating $\tilde{\Upsilon}_1$ and going back to the original fields 
without tilde, one obtains the Lagrangian in terms of $\Upsilon_0$, the Lagrange 
multiplier which is denoted as $Y_0$, $\hat{V}_{-1}, \hat{V}_0$ and $\hat{V}_1$.

Now we consider the $\cN=3$ gauge interacting Lagrangian in terms of 
the $\mathcal{N} = 2$ superfield language. The charged hypermultiplet consists 
of two chiral superfields $(S, T)$ and they couple with the non-Abelian vector 
multiplet ($V, \Phi, \bar{\Phi}$). The Lagrangian is given by
\begin{align}
\mathcal{L} = \int \! d^4 \theta \ 
\left(
\bar{S} e^{V} S + T e^{-V} \bar{T}
\right)
+ 
\left[
2 \int \! d^2 \theta \ T \Phi S - 2 \int \! d^2 \bar{\theta} \ \bar{T}
 \bar{\Phi} \bar{S}
\right].
\end{align}
We can directly compare this with the Lagrangian in terms of $\Upsilon_0, Y_0$,
$\hat{V}_{-1}, \hat{V}_0$ and $\hat{V}_1$coming from (\ref{eq:hyper}). 
In \cite{ArSa2} we found that $S = \Upsilon_0$, $T = \bar{\mathcal{D}}^2
\bar{\tilde{Y}}_0 e^{\hat{V}_0}$ are chiral superfields and 
$\bar{\tilde{Y}}_0 = \bar{Y}_0 e^{\hat{V}_{-}}$ and
\begin{align}
V =& \hat{V}_0, \notag \\
\Phi =& \frac{1}{8} e^{- \hat{V}_0}  
\left(
- 2 D^{12 \alpha} \Gamma^{(-)}_{0 \alpha} + \bar{\mathbb{D}}^{\alpha}
 \Gamma^{(-)}_{-1 \alpha} + \{\Gamma^{(-) \alpha}_{-1},
 \Gamma^{(-)}_{0 \alpha} \}
\right) e^{\hat{V}_0}, \notag \\
\bar{\Phi} =& 
\frac{1}{8}
\left(
- 2 D^{12 \alpha} \Gamma^{(-)}_{-2 \alpha} -
 \mathbb{D}^{\alpha} \Gamma^{(-)}_{-1 \alpha} + \{\Gamma^{(-)
 \alpha}_{-2}, \Gamma^{(-)}_{-1 \alpha} \}
\right).
\label{eq:vector_tropical}
\end{align}
In \cite{ArSa2} we have checked that $\Phi$ ($\bar{\Phi}$) defined in the above 
actually satisfies the (anti)chirality condition
$\bar{\mathbb{D}}_{\alpha} \Phi = \mathbb{D}_{\alpha} \bar{\Phi} = 0$.
We note that in the Abelian limit, we reproduce the correct expression found in \cite{ArSa1}
\begin{align}
V = V_0, \quad \Phi = - \frac{1}{8} \bar{\mathbb{D}}^2 V_1, \quad 
\bar{\Phi} = - \frac{1}{8} \mathbb{D}^2 V_{-1}.
\end{align}

In the next section, we propose the supersymmetric $\cN=3$ and $\cN=4$ 
Chern-Simons Lagrangians by utilizing the above results.

\section{Chern-Simons Lagrangian}
In this section, we construct 
a Lagrangian for the non-Abelian Chern-Simons theory in the projective superspace. 
We also generalize the result to the $\mathcal{N} = 4$ case.
The three-dimensional $\mathcal{N} = 3$ Chern-Simons Lagrangian 
in the $\mathcal{N} = 2$ superfield formalism is \cite{Iv}
\begin{align}
\mathcal{L}^{\mathcal{N} = 3}_{\mathrm{CS}} 
=& 
- \frac{ik}{4 \pi} \int \! d^4 \theta 
\int^1_0 \! dt \ \mathrm{Tr}
\left[
V \bar{\mathbb{D}}^{\alpha} (e^{t V} \mathbb{D}_{\alpha} e^{- t
V})
\right]
- \frac{k}{4\pi} \int \! d^2 \theta \ \mathrm{Tr} \Phi^2
+ \frac{k}{4\pi} \int \! d^2 \bar{\theta} \ \mathrm{Tr} \bar{\Phi}^2.
\label{eq:N2N3CS}
\end{align}
Here $k$ is the Chern-Simons level and $t$ is an auxiliary integration
variable. 
The first term can be rewritten as 
\begin{align}
- \frac{ik}{4 \pi} \int \! d^4 \theta 
\int^1_0 \! dt \ \mathrm{Tr}
\left[
V \bar{\mathbb{D}}^{\alpha} (e^{t V} \mathbb{D}_{\alpha} e^{- t
V})
\right]
= 
\frac{ik}{4 \pi} \int \! d^4 \theta 
 \int^1_0 \! dt \ \mathrm{Tr}
 \left[
 V \mathbb{D}^{\alpha} (e^{- t V} 
 \bar{\mathbb{D}}_{\alpha}
  e^{ t V})
 \right],
\label{eq:N2N3CS2}
\end{align}
where we have used the fact that $A(MBM^{-1}) = MB(M^{-1} A M) M^{-1}$
for variables $A,B$ and $M$ that satisfy $\{A,B\} = 0$.

We first start from the $\mathcal{N} = 3$ Abelian Chern-Simons theory.
When the gauge group is Abelian, the Lagrangian \eqref{eq:N2N3CS} 
is reduced to the following form:
\begin{align}
\mathcal{L} = \frac{ik}{8 \pi} \int \! d^4 \theta \ V
 \bar{\mathbb{D}}^{\alpha} \mathbb{D}_{\alpha} V - \frac{k}{4 \pi} \int
 \! d^2 \theta \ \Phi^2 + \frac{k}{4\pi} \int \! d^2 \bar{\theta} \
 \bar{\Phi}^2.
\label{eq:Abelian_CS}
\end{align}
We have constructed the Lagrangian in the projective superspace 
that reproduces the expression \eqref{eq:Abelian_CS}. 
The Lagrangian in the $\mathcal{N} = 3$ projective superspace is \cite{ArSa1}
\begin{align}
\mathcal{L} = \frac{k}{8 \pi} \oint_{\gamma} \! \frac{d \zeta}{2 \pi i
 \zeta} \int \! d^4 \theta \ 
\mathcal{V}^{[0]} G^{[2]},
\label{eq:N3CS_abelian}
\end{align}
where the gauge invariant $\mathcal{O} (-1,1)$ multiplet $G^{[2]}$ with
weight 2 is a function of the tropical multiplet $\mathcal{V}^{[0]}$ and 
is expanded as 
\begin{align}
G^{[2]} = \frac{i}{\zeta} \Phi_0 + L + i \zeta \bar{\Phi}_0.
\end{align}
Each component should satisfy the 
projective superfield condition $D^{[2]}_{\alpha} G^{[2]} = 0$, namely, 
\begin{align}
\mathbb{D}_{\alpha} \bar{\Phi}_0 = 0, \qquad 
\mathbb{D}^2 L = \bar{\mathbb{D}}^2 L = 0, \qquad 
\bar{\mathbb{D}}_{\alpha} \Phi_0 = 0.
\label{eq:fs_projective}
\end{align}
Taking account of the gauge invariance of $G^{[2]}$,
these constraints are solved by 
\begin{align}
L = i \bar{\mathbb{D}}^{\alpha} \mathbb{D}_{\alpha} V_0, \quad 
\Phi_0 = - \frac{1}{8} \bar{\mathbb{D}}^2 V_1, \quad 
\bar{\Phi}_0 = - \frac{1}{8} \mathbb{D}^2 V_{-1}.
\label{eq:abelian_O}
\end{align}
This is a gauge-fixed form of the relation found in \cite{KuLiTa}.
It is easy to check that $\Phi_0$ ($\bar{\Phi}_0$) is 
the (anti)chiral superfield appearing in \eqref{eq:Abelian_CS} by substituting 
(\ref{eq:abelian_O}) into (\ref{eq:N3CS_abelian}).

Now we construct a Lagrangian for the non-Abelian Chern-Simons theory 
in the $\mathcal{N} = 3$ projective superspace.
For non-Abelian gauge groups, the embedding of the tropical multiplet 
into the $\mathcal{O} (-1,1)$ multiplet \eqref{eq:abelian_O} becomes non-linear.
We assume that the Lagrangian is given by the product of a 
function of $\mathcal{V}^{[0]}$ and an $\mathcal{O} (-1,1)$ multiplet $G^{[2]}$
even for non-Abelian gauge groups.
Since the Lagrangian should reproduce the result \eqref{eq:N3CS_abelian}
in the Abelian limit, we propose the following non-Abelian Chern-Simons 
Lagrangian in the $\mathcal{N} = 3$ projective superspace:
\begin{eqnarray}
\mathcal{L} 
= \frac{k}{8 \pi} \oint_{\gamma} \! \frac{d \zeta}{2 \pi i \zeta} 
\int \! d^4 \theta \
\mathrm{Tr} 
\left[
f (\mathcal{V}^{[0]}) G^{[2]} \right],
\label{eq:N3CS}
\end{eqnarray}
where $f$ is a projective superfield with weight 0 which is 
a function of the tropical multiplet $\mathcal{V}^{[0]}$ and
$G^{[2]} = \frac{i}{\zeta} \Phi_0 + L + i \zeta \bar{\Phi}_0$ 
is an $\mathcal{O} (-1,1)$ multiplet with weight 2 which satisfies the 
constraints \eqref{eq:fs_projective}.
The function $f$ should satisfy $f(\mathcal{V}^{[0]}) \to
\mathcal{V}^{[0]}$ in the Abelian limit.
Although the gauge invariance of the action is not manifest, the
component expression of the action will ensure it.
In the Lindstr\"om-Ro\v{c}ek gauge, the $\zeta$ expansion of the function $f$
is generically given by 
\begin{eqnarray}
f (\mathcal{V}^{[0]}) = \frac{1}{\zeta} f_{-1}(z) + f_0(z) + \zeta
 f_1(z) + \cdots, 
\end{eqnarray}
where $\cdots$ are terms that contain $\zeta^n, (n \neq 0, \pm 1)$ which
 are irrelevant in the Lagrangian \eqref{eq:N3CS} and vanish in the
 Abelian limit.
Performing the $\zeta$ integration, the Lagrangian 
is reduced to that in the $\mathcal{N} = 2$ superspace,
\begin{eqnarray}
\mathcal{L} 
= \frac{ik}{8 \pi} \int \! d^4 \theta 
\mathrm{Tr}
\left[
 f_1 \Phi_0 -i f_0 L + f_{-1} \bar{\Phi}_0
\right].
\label{eq:N3CS_decomp}
\end{eqnarray}
We look for the explicit forms of the components $(f_{-1}, f_0, f_1)$ and
$(\Phi_0, L, \bar{\Phi}_0)$ that reproduce the action \eqref{eq:N2N3CS} in
the $\mathcal{N} = 2$ superspace. 
First, we identify $\Phi_0, \bar{\Phi}_0$ with the non-Abelian 
adjoint (anti)chiral superfields $\Phi, \bar{\Phi}$ defined in
\eqref{eq:vector_tropical} (with extra overall factors $\pm i$):
\begin{eqnarray}
&&\Phi_0 = i \Phi = \frac{i}{8} e^{- \hat{V}_0}  
\left(
- 2 D^{12 \alpha} \Gamma^{(-)}_{0 \alpha} + \bar{\mathbb{D}}^{\alpha}
 \Gamma^{(-)}_{-1 \alpha} + \{\Gamma^{(-) \alpha}_{-1},
 \Gamma^{(-)}_{0 \alpha} \}
\right) e^{\hat{V}_0}, \label{Phi-1} \\
&&\bar{\Phi}_0 = - i \bar{\Phi} = 
- \frac{i}{8}
\left(
- 2 D^{12 \alpha} \Gamma^{(-)}_{-2 \alpha} -
 \mathbb{D}^{\alpha} \Gamma^{(-)}_{-1 \alpha} + \{\Gamma^{(-)
 \alpha}_{-2}, \Gamma^{(-)}_{-1 \alpha} \}
\right). \label{Phi-2}
\end{eqnarray}
Then, using the constraints \eqref{eq:tropical_constraint}, we 
can show that the above definition satisfies a part of the projective superfield constraints of 
$G^{[2]}$, $\mathbb{D}_{\alpha} \bar{\Phi}_0 =\bar{\mathbb{D}}_{\alpha} \Phi_0 = 0$ \cite{ArSa2}.
Next, we rewrite parts of the D-terms in the Lagrangian
\eqref{eq:N3CS_decomp} to F-terms:
\begin{eqnarray}
\mathcal{L}
= \frac{ik}{8 \pi} \int \! d^4 \theta \
\mathrm{Tr} 
\left[
-i f_0 L 
\right]
- \frac{k}{8 \pi} 
\int \! d^2 \theta \
\mathrm{Tr}
\left[
- \frac{1}{4} \bar{\mathbb{D}}^2 f_1 \Phi
\right]
+ \frac{k}{8 \pi}
\int \! d^2 \bar{\theta} \
\mathrm{Tr}
\left[
- \frac{1}{4} \mathbb{D}^2 f_{-1} \bar{\Phi}
\right],
\label{eq:N3CS_decomp_DF}
\end{eqnarray}
where we have used the fact that $\bar{\mathbb{D}}_{\alpha} \Phi =
\mathbb{D}_{\alpha} \bar{\Phi} = 0$ and dropped the total derivative
terms. Comparing the first term in \eqref{eq:N3CS_decomp_DF} with the
D-term in the Lagrangian \eqref{eq:N2N3CS},
we determine the components $L$ as 
\begin{align}
L = - 2 \bar{\mathbb{D}}^{\alpha} 
\int^1_0 \! d t 
\left(
e^{t \hat{V}_0} \mathbb{D}_{\alpha} e^{- t \hat{V}_0}
\right),
\label{eq:L1}
\end{align}
and $f_0$ as 
\begin{align}
f_0 = \hat{V}_0.
\end{align}
When we employ the expression \eqref{eq:N2N3CS2} instead of \eqref{eq:N2N3CS}, we have 
\begin{align}
L =  2 \mathbb{D}^{\alpha}
\int^1_0 \! dt 
\left(
e^{- t \hat{V}_0} \bar{\mathbb{D}}_{\alpha}  e^{ t \hat{V}_0}
\right).
\label{eq:L2}
\end{align}
The two expressions of $L$, \eqref{eq:L1} and \eqref{eq:L2}, are physically
equivalent.
Since $L$ is a component of the $\mathcal{O} (-1,1)$ multiplet
$G^{[2]}$, it should satisfy the constraint $\mathbb{D}^2 L =
\bar{\mathbb{D}}^2 L = 0$. We examine this condition in the following.
Because the calculations for the expressions \eqref{eq:L1} and \eqref{eq:L2}
are essentially the same, we focus on the expression \eqref{eq:L1}
\footnote{When we employ the other expression \eqref{eq:L2}, the following calculations hold
if $\mathbb{D}_{\alpha}$ and $\bar{\mathbb{D}}_{\alpha}$ are interchanged.}.
The expression \eqref{eq:L1} is manifestly $\bar{\mathbb{D}}$-exact
form.
Therefore we find that the first constraint is satisfied trivially:
\begin{equation}
\bar{\mathbb{D}}^2 L = 0.
\end{equation}
On the other hand, the second condition $\mathbb{D}^2 L = 0$ is not
manifest. 
We examine the second condition by perturbative calculations.
We concentrate on the next leading order in $V$ where the non-Abelian property begins to appear
for the first time
\footnote{
The leading order $\mathcal{O} (V)$ corresponds to the Abelian case.
Here $V$ represents
the components $V_{-1}, V_0, V_1$ in the tropical multiplet $\mathcal{V}^{[0]}$.}.
Up to $\mathcal{O} (V^3)$, we have
\begin{align}
& \mathbb{D}^2 
\left[
- 2 \bar{\mathbb{D}}^{\alpha} 
\left(
e^{t \hat{V}_0} \mathbb{D}_{\alpha} e^{- \hat{V}_0}
\right)
\right] 
\notag \\
& = 
- t^2 
\left(
[\mathbb{D}^2 \mathbb{D}^{\alpha} \hat{V}_0, \hat{V}_0] 
+
\{
\mathbb{D}^{\beta} \mathbb{D}^{\alpha} \bar{\mathbb{D}}_{\alpha}
 \hat{V}_0, \mathbb{D}_{\beta} \hat{V}_0
\}
+
\{
\mathbb{D}^{\beta} \mathbb{D}^{\alpha} \bar{\mathbb{D}}_{\alpha}
 \hat{V}_0, \mathbb{D}_{\beta} \hat{V}_0
\}
+
[\mathbb{D}^{\alpha} \bar{\mathbb{D}}_{\alpha} \hat{V}_0, \mathbb{D}^2
 \hat{V}_0]
\right. 
\notag \\
& \left.
- \{
\mathbb{D}^2 \mathbb{D}^{\alpha} \hat{V}_0, \bar{\mathbb{D}}_{\alpha} \hat{V}_0
\}
+
[\mathbb{D}^{\beta} \mathbb{D}^{\alpha} \hat{V}_0, \mathbb{D}_{\beta}
 \bar{\mathbb{D}}_{\alpha} \hat{V}_0]
-
[
\mathbb{D}^{\beta} \mathbb{D}^{\alpha} \hat{V}_0, \mathbb{D}_{\beta}
 \bar{\mathbb{D}}_{\alpha} \hat{V}_0
]
+
\{
\mathbb{D}^{\alpha} \hat{V}_0, \mathbb{D}^2 \bar{\mathbb{D}}_{\alpha} \hat{V}_0
\}
\right)
+ \mathcal{O} (V^3).
\end{align}
For non-Abelian gauge groups, the $\mathcal{N} = 2$ vector superfield
$V = \hat{V}_0$ is perturbatively expressed as 
\begin{align}
\hat{V}_0 = V_0 + \frac{1}{2} [V_1, V_{-1}] + \mathcal{O}(V^3).
\end{align}
Using the constraints for the tropical multiplet $\mathcal{V}^{[0]}$, 
we find 
\begin{align}
 \mathbb{D}^2 
\left[
- 2 \bar{\mathbb{D}}^{\alpha} 
\left(
e^{t \hat{V}_0} \mathbb{D}_{\alpha} e^{-t \hat{V}_0}
\right)
\right] 
=& - t^2 
\left(
-
\{
\mathbb{D}^{\alpha} V_0, \mathbb{D}^2 \bar{\mathbb{D}}_{\alpha} V_0 
\}
+ 
\{
\mathbb{D}^{\alpha} V_0, \mathbb{D}^2 \bar{\mathbb{D}}_{\alpha} V_0
\}
\right)
+ \mathcal{O} (V^3)
\notag \\
=& 0 + \mathcal{O} (V^3).
\end{align}
Then up to $\mathcal{O} (V^3)$, we find that the expression
\eqref{eq:L1} satisfies the constraints $\mathbb{D}^2 L =
\bar{\mathbb{D}}^2 L = 0$. 
Therefore, all the components $(\Phi_0, L, \bar{\Phi}_0)$ 
are correctly embedded into the $\mathcal{O} (-1,1)$ multiplet $G^{[2]}$.

Finally, we look for expressions of the functions $f_1, f_{-1}$.
Comparing the second and the third terms in 
\eqref{eq:N3CS_decomp_DF} with the F-terms in the Lagrangian \eqref{eq:N2N3CS},
we find the following relations, 
\begin{eqnarray}
\bar{\mathbb{D}}^2 f_1 = - 8 \Phi, \qquad \mathbb{D}^2 f_{-1} = - 8 \bar{\Phi}.
\end{eqnarray}
We need to solve $f_{-1}, f_1$ in the above relations. 
Since the anti-chiral superfield $\bar{\Phi}$ (\ref{Phi-2}) is shown to be 
$\mathbb{D}^2$-exact form \cite{ArSa2}
\begin{eqnarray}
\bar{\Phi} = - \frac{1}{8} \mathbb{D}^2 \hat{V}_{-1}, 
\end{eqnarray}
the function $f_{-1}$ is determined to be 
\begin{eqnarray}
f_{-1} =  \hat{V}_{-1}.
\end{eqnarray}
On the other hand, the chiral superfield $\Phi$ (\ref{Phi-1}) is not manifestly 
$\bar{\mathbb{D}}^2$-exact. Again, we calculate the function $f_{1}$ by
perturbation in $V$. 
From (\ref{Phi-1}) we have 
\begin{eqnarray}
 \Phi  
= - \frac{1}{8}\bar{\mathbb{D}}^2 V_1 + \frac{1}{16} [V_0, \bar{\mathbb{D}}^2 V_1] 
- \frac{1}{8} \{ \bar{\mathbb{D}}^{\alpha} V_1, \bar{\mathbb{D}}_{\alpha} V_0 \} +
\mathcal{O} (V^3).
\end{eqnarray}
Using the projective superspace constraints of the tropical multiplet, we find that 
the chiral superfield is rewritten as the $\mathbb{D}^2$-exact form up to 
$\mathcal{O} (V^3)$ calculation,
\begin{eqnarray}
\Phi = 
\frac{1}{8}
\bar{\mathbb{D}}^2 
\left(
- V_1 + \frac{1}{2} [V_0, V_1]
\right) + \mathcal{O} (V^3).
\end{eqnarray}
Then, the function $f_{1}$ is determined to be 
\begin{equation}
f_{1} =  
\left(
V_1 + \frac{1}{2} [V_1, V_0]
\right) + \mathcal{O} (V^3).
\label{eq:f1}
\end{equation}
Therefore all the functions $f_{\pm1}, f_0$ have been determined.
We stress that the expression \eqref{eq:f1} is 
nothing but the first two terms in the perturbative expansion of
$\hat{V}_1$ in \eqref{eq:potential}. 
Although our calculations are limited to the perturbative regime, 
this result suggests that $f_1$ is naturally given by $f_1 = \hat{V}_1$ for the
full order in $V$.
All the expressions $f_{\pm 1} = \hat{V}_{\pm 1}, f_0 = \hat{V}_0$ have
the correct Abelian limit $f_{\pm 1} \to V_{\pm 1}, f_0 \to V_0$.
In summary we have found explicit forms of the functions $f_{-1}, f_0, f_1$ and the
$\mathcal{O} (-1,1)$ multiplet in the Lagrangian \eqref{eq:N3CS} in the
$\mathcal{N} = 3$ projective superspace.
We note that only $\hat{V}_{-1}$, $\hat{V}_0$, $\hat{V}_1$,
$\hat{V}_2$ in the decomposition \eqref{eq:decomp} 
appear in the Lagrangian.

We generalize this result to the $\mathcal{N} = 4$ model.
For the $\mathcal{N} = 4$ theory, a pair of the projective multiplets
associated with the two $\mathbb{C}P^1$s (see Appendix B) is introduced.
We propose the following Lagrangian for the 
$\mathcal{N} = 4$ generalization of the $\mathcal{N} = 3$ model \eqref{eq:N3CS}:
\begin{align}
\mathcal{L} = 
\frac{k}{8 \pi} \oint \! \frac{d \zeta_L}{2 \pi i \zeta_L} 
d^4 \theta \ \mathrm{Tr} [f (\mathcal{V}^{[0]}_L) G^{[2]}_L]
+ 
\frac{k}{8 \pi} \oint \! \frac{d \zeta_R}{2 \pi i \zeta_R} 
d^4 \theta \ \mathrm{Tr} [f (\mathcal{V}^{[0]}_R) G^{[2]}_R],
\end{align}
where the function $f$ is the same one found in the $\mathcal{N} = 3$
model. The $\mathcal{O} (-1,1)$ multiplet of the left sector $G^{[2]}_L$
is the function of the right tropical multiplet 
$\mathcal{V}^{[0]}_R$ and vice versa:
\begin{align}
G^{[2]}_{L,R} = \frac{i}{\zeta_{L,R}} \Phi_{R,L} + L_{R,L} 
+ i \zeta_{L,R} \bar{\Phi}_{R,L}.
\end{align}
In order to be consistent with the Abelian limit \cite{ArSa1}, 
we take the each component in $G^{[2]}_{L,R}$ as the same one in the $\mathcal{N} =
3$ case. 
Here the superfields $(V_L, \Phi_L, \bar{\Phi}_L)$, 
$(V_R, \Phi_R, \bar{\Phi}_R)$ are defined by the components in 
$\mathcal{V}^{[0]}_L$ and $\mathcal{V}^{[0]}_R$ as in the $\mathcal{N} =3$ case. 
Performing the $\zeta_L$, $\zeta_R$ integration, we find  
\begin{align}
\mathcal{L} =&
- \frac{ik}{4 \pi} \int \! d^4 \theta 
\int^1_0 \! dt \ \mathrm{Tr}
\left[
V_L \bar{\mathbb{D}}^{\alpha} (e^{t V_R} \mathbb{D}_{\alpha} e^{- t
V_R})
\right]
- \frac{k}{4\pi} \int \! d^2 \theta \ \mathrm{Tr} \Phi_L \Phi_R
+ \frac{k}{4\pi} \int \! d^2 \bar{\theta} \ \mathrm{Tr} \bar{\Phi}_L
 \bar{\Phi}_R
\notag \\
& + 
\left(
L \leftrightarrow R
\right).
\label{eq:N4CS}
\end{align}
This Lagrangian contains two gauge fields and 
mixing interactions between the left and right multiplets.
This kind of theory is known as the BF-theory \cite{Ho, BiBlRaTh}. 
The $\mathcal{N} = 4$ supersymmetric
BF-theory is discussed in the harmonic superspace formalism \cite{Zu}.

\section{Conclusion and discussions}
In this letter we have studied the $\mathcal{N} = 3$ and $\mathcal{N} = 4$
non-Abelian Chern-Simons actions in the three-dimensional projective
superspaces. We work in the projective superspaces where the
superconformal symmetry is manifest.
The $\mathcal{N} = 3$ and $\mathcal{N} = 4$ vector multiplets are defined
by the tropical multiplet $\mathcal{V}^{[0]}$ with weight 0.
The relations among the component superfields $(V_{-1}, V_0, V_1)$ in $\mathcal{V}^{[0]}$ and
the vector multiplet $(V, \Phi, \bar{\Phi})$ in the $\mathcal{N}
= 2$ superspace are quite non-linear for non-Abelian gauge groups.
The explicit relations among them are found
in our previous paper \cite{ArSa2}.

In this letter, using the explicit relations of the component
superfields, we propose the Lagrangian \eqref{eq:N3CS} for the 
superconformal non-Abelian Chern-Simons theory in the $\mathcal{N} = 3$ projective superspace. 
Although we have a little principle to determine the function $f$
of the tropical multiplet $\mathcal{V}^{[0]}$, we have found  
the explicit form of the function by the help of the action in the $\mathcal{N} =
2$ superspace. 
The $\zeta^{\pm 1}, \zeta^0$ components of the function $f$ consist of $\hat{V}_{\pm 1},
\hat{V}_0$ which appeared in the decomposition \eqref{eq:decomp} of the non-Abelian tropical
multiplet $\mathcal{V}^{[0]}$.
We also found the explicit embedding of the non-Abelian tropical
multiplet $\mathcal{V}^{[0]}$ into the $\mathcal{O} (-1,1)$
multiplet $G^{[2]}$ with weight 2. The $\mathcal{O} (-1,1)$ multiplet plays the role
of the gauge field strength associated with the gauge potential
$\mathcal{V}^{[0]}$. The Lagrangian \eqref{eq:N3CS} has the correct
Abelian limit \cite{ArSa1}.
We demonstrated that the proposed Lagrangian \eqref{eq:N3CS} successfully
reproduces the $\mathcal{N} = 3$ non-Abelian Chern-Simons Lagrangian in the
$\mathcal{N} = 2$ superspace \cite{Iv}.
We also discussed the $\mathcal{N} = 4$ generalization of our
Lagrangian.
We stress that although our calculations are based on the perturbation, they are
not trivial even in the next leading order in $V$.
Moreover, the very suggestive expression \eqref{eq:f1} implies that our
analysis holds true even for the full order in $V$.

We found the functions $f$ and $G^{[2]}$ in the language of
the component superfields of $\mathcal{V}^{[0]}$ in this letter.
For an Abelian gauge group, the $\mathcal{O} (-1,1)$ multiplet $G^{[2]}$ is a linear function of
$\mathcal{V}^{[0]}$ \cite{ArSa1, KuLiTa}. 
Since for a non-Abelian case, this would become non-linear and complicated, 
it is challenging to write down the Lagrangian in terms of the
projective superfield $\mathcal{V}^{[0]}$.

For an application of the present formalism, it is interesting to write down the
$\mathcal{N} = 6$ ABJM action \cite{AhBeJaMa} in the 
projective superspaces \footnote{The $\mathcal{N} = 6$ ABJM action has been
constructed in the harmonic superspace formalism \cite{ABJM-HSS}. 
The Chern-Simons action based on the 3-algebra has been discussed in \cite{ChDoSa}.
}. 
The gauge field part of the ABJM model is the $U(N) \times U(N)$ Chern-Simons
model with opposite Chern-Simons level ($k, -k$).
We can easily construct the $\mathcal{N} = 6$ ABJM action in the
$\mathcal{N} = 3$ projective superspace.
However, in the $\mathcal{N} = 4$ projective superspace, 
the first term in the Lagrangian \eqref{eq:N4CS} is the mixing term
of $V_L$ and $V_R$. Then it is not the standard Chern-Simons term discussed
in \cite{Iv} but is the BF-theory.
At least in the component level in the 
Abelian limit, we found that the first term
is rewritten as the sum of the two Chern-Simons terms 
constructed by the two vector superfields $V$ and $V'$ with opposite
Chern-Simons level ($k,-k$).
Here $V = \frac{1}{\sqrt{2}} (V_L + V_R)$ and $V' = \frac{1}{\sqrt{2}}
(V_L - V_R)$.
For the non-Abelian case, $V$ and $V'$ would become highly non-linear functions of $V_L$ and $V_R$. 
Moreover, in order to incorporate with the bi-fundamental matters which couples
to left and right parts of the gauge potentials, one may need the hybrid
projective multiplet \cite{KuLiTa}. 
Non-Abelian gauge interactions of the hybrid projective multiplet in the
$\mathcal{N} = 4$ projective superspace is also interesting.
We will come back to these issues in the future works.

\subsection*{Acknowledgments}
The work of M. A. is supported by Grant-in-Aid for 
Scientific Research from the Ministry of Education, Culture, 
Sports, Science and Technology, Japan (No.25400280)
and in part by the Research Program MSM6840770029 and 
by the project of International Cooperation ATLAS-CERN of 
the Ministry of Education, Youth and Sports of the Czech Republic.
The work of S.~S. is supported in part by Sasakawa Scientific 
Research Grant from The Japan Science Society and Kitasato University
Research Grant for Young Researchers.

\begin{appendix}
\section{
Conventions and notations of ordinary superspaces in three dimensions
}\label{appendixA}
In this section, we provide the basic conventions 
and notations of the $\mathcal{N} =2$, $\mathcal{N} = 3$ and 
$\mathcal{N} = 4$ superspaces in three dimensions.
The three-dimensional metric is given by $\eta_{mn} = \mathrm{diag} (-1,+1,+1)$.
The three-dimensional $\mathcal{N} = 2$ superspace is represented by the 
coordinates $(x^{m}, \theta^{\alpha}, \bar{\theta}^{\alpha})$ where 
$\theta, \bar{\theta}$ are two component spinors. 
The index $\alpha=1,2$ is associated with the $SO(1,2) \sim SL(2,\mathbb{R})$ 
Lorentz spinors.
The spinor indices are raised and lowered by the anti-symmetric epsilon
symbol $\varepsilon^{12} = - \varepsilon_{12} = 1$. 
The gamma matrices which satisfy the Clifford algebra 
$\{\gamma^{m}, \gamma^{n} \} = 2 \eta^{mn}$ 
are defined by $(\gamma^{m})_{\alpha} {}^{\beta} = (i \tau^2, \tau^1 ,\tau^3)$.
Here $\tau^I \ (I=1,2,3)$ are the Pauli matrices and 
$I = 1,2, 3$ is the vector index of the $SO(3)_R \sim SU(2)_R$ R-symmetry.
The supercovariant derivatives in the $\mathcal{N} = 2$ superspace 
are defined by 
\begin{eqnarray}
\begin{aligned}
& \mathbb{D}_{\alpha} = \partial_{\alpha} + i (\gamma^{m} \bar{\theta})_{\alpha} 
\partial_{m}, \qquad 
\bar{\mathbb{D}}_{\alpha} = - \bar{\partial}_{\alpha} - i (\theta 
\gamma^{m})_{\alpha} \partial_{m}, \label{cov1} \\
& \{\mathbb{D}_{\alpha}, \bar{\mathbb{D}}_{\beta} \} = - 2 i
 \gamma^{m}_{\alpha \beta} \partial_{m}, \quad \{ \mathbb{D}_{\alpha},
 \mathbb{D}_{\beta} \} = \{\bar{\mathbb{D}}_{\alpha},
 \bar{\mathbb{D}}_{\beta} \} = 0.
\end{aligned}
\end{eqnarray}
The Grassmann measure of integration in the $\mathcal{N} = 2$ superspace
is defined by 
\begin{eqnarray}
d^2 \theta = - \frac{1}{4} d \theta^{\alpha} d \theta_{\alpha}, \quad 
d^2 \bar{\theta} = - \frac{1}{4} d \bar{\theta}^{\alpha} d 
\bar{\theta}_{\alpha}, \quad d^4 \theta = d^2 \theta d^2 \bar{\theta}.
\end{eqnarray}
They are normalized such that,
\begin{eqnarray}
\int \! d^2 \theta \ \theta^2 = 1, \quad \int \! d^2 \bar{\theta} \ 
\bar{\theta}^2=1, \quad \int \! d^4 \theta \ \theta^2 \bar{\theta}^2 = 1.
\end{eqnarray}
For an $\mathcal{N} = 2$ superfield 
$F (x,\theta, \bar{\theta})$, 
the following relation holds within the spacetime integration, 
\begin{eqnarray}
\int \! d^4 \theta \ 
F (x, \theta, \bar{\theta}) = \left. \frac{1}{16} 
(\mathbb{D}^2 \bar{\mathbb{D}}^2 
F (x, \theta, \bar{\theta})
) 
\right|_{\theta = \bar{\theta} = 0}.
\end{eqnarray}
The chiral and anti-chiral coordinates are defined by 
\begin{eqnarray}
x^{m}_L = x^{m} + i \theta \gamma^{m} \bar{\theta}, \qquad 
x^{m}_R = x^{m} - i \theta \gamma^{m} \bar{\theta}.
\end{eqnarray}
\end{appendix}
The $\mathcal{N} = 3$ superspace coordinates are defined by 
$z=(x^{m},\theta^{\alpha}_{ij})$
where $i=1,2$ is the $SU(2)_R$ R-symmetry spinor index
and the Grassmann coordinate satisfies the reality condition
$\overline{\theta^\alpha_{ij}}=\theta^{\alpha ij}$.
%
The $SU(2)_R$ spinor indices and the $SO(3)_R$ vector indices are intertwined by the relation
$\theta^{\alpha}_{ij} = (\tau_I)_{ij} \theta^{\alpha}_I$.
The $SU(2)_R$ indices are raised and 
lowered by the anti-symmetric symbols $\varepsilon^{ij}, \varepsilon_{ij}$.
The supercovariant derivatives in the $\mathcal{N} = 3$ superspace are defined by 
\begin{eqnarray}
\begin{aligned}
& D^{ij}_{\alpha} = \frac{\partial}{\partial \theta^{\alpha}_{ij}} 
+ i \theta^{\beta}_{ij} \partial_{\alpha \beta},\qquad 
\partial_{\alpha\beta} = \gamma^m_{\alpha\beta}\partial_m, \\
& \{D^{ij}_{\alpha}, D^{kl}_{\beta} \} = - 2 i \varepsilon^{i(k}
 \varepsilon^{l)j} \partial_{\alpha \beta}.
\end{aligned}
\end{eqnarray}
The $\mathcal{N} = 4$ superspace coordinates are defined by 
$z' = (x^m, \theta^{\alpha}_{i \bar{j}})$
where $i = 1,2$ and $\bar{j} = 1,2$ are indices for the 
$SU(2)_L \times SU(2)_R$ subgroup of $SO(4)_R$ R-symmetry
and the Grassmann coordinate satisfies the reality condition
$\overline{\theta^\alpha_{i\bar{j}}}=\theta^{\alpha i\bar{j}}$.
The supercovariant derivatives in the $\mathcal{N} = 4$ superspace are
defined by 
\begin{eqnarray}
\begin{aligned}
& D^{i \bar{j}}_{\alpha} = \frac{\partial}{\partial \theta^{\alpha}_{i
 \bar{j}}} + i \theta^{\beta}_{i \bar{j}} \partial_{\alpha \beta}, \\
& \{D^{i \bar{j}}_{\alpha}, D^{k \bar{l}}_{\beta} \} = 
 2 i \varepsilon^{ik} \varepsilon^{\bar{j} \bar{l}} \partial_{\alpha
 \beta}.
\end{aligned}
\end{eqnarray}
We use the following relations among the $\mathcal{N} = 2$, 
$\mathcal{N} = 3$ and $\mathcal{N} = 4$ superspaces \cite{KuPaTaUn}:
\begin{eqnarray}
& & \theta^{\alpha} = \theta^{\alpha}_{11} = \theta^{\alpha}_{1\bar{1}},
 \quad 
\bar{\theta}^{\alpha} = \theta^{\alpha}_{22} =
\theta^{\alpha}_{2\bar{2}}, \\
& & \mathbb{D}_{\alpha} = D^{11}_{\alpha} = D^{1 \bar{1}}_{\alpha},
 \quad 
\bar{\mathbb{D}}_{\alpha} = - D^{22}_{\alpha} = - D^{2\bar{2}}_{\alpha}.
\end{eqnarray}

\section{Projective superspace formalisms}
In this section, we summarize conventions and notations of the projective
superspaces in three dimensions. For more detail, see \cite{KuPaTaUn, ArSa1, ArSa2}. 

\subsection{$\mathcal{N} = 3$ projective superspace}
We introduce the $SU(2)_R$ complex isospinors $v^i, u^i$ ($i = 1,2$) 
which satisfy the following completeness relation, 
\begin{eqnarray}
\delta^i {}_j = \frac{1}{(v,u)} (v^i u_j - v_j u^i), \quad
(v,u) \equiv v^i u_i \not= 0.
\label{completeness}
\end{eqnarray}
The supercovariant derivative in the projective superspace is defined as 
\begin{eqnarray}
D^{(2)}_{\alpha} = v_i v_j D^{ij}_{\alpha}, \quad
D^{(0)}_{\alpha} = \frac{1}{(v,u)} v_i u_j D^{ij}_{\alpha}, \quad 
D^{(-2)}_{\alpha} = \frac{1}{(v,u)^2} u_i u_j D^{ij}_{\alpha}.
\end{eqnarray}
A projective superfield $Q^{(n)}$ with weight $n$ is defined by 
\begin{eqnarray}
D^{(2)}_{\alpha} Q^{(n)} = 0, \qquad 
Q^{(n)} (z, cv) = c^n Q^{(n)} (z, v), \quad c \in \mathbb{C}^{*}.
\end{eqnarray}
The $\mathcal{N} = 3$ superconformal invariant action is 
\begin{eqnarray}
S = 
 \frac{1}{8\pi} \oint_{\gamma} \!
(v, dv) 
 \int \! d^3 x 
\ (D^{(-2)})^2 (D^{(0)})^2 \left. \mathcal{L}^{(2)} (z,v)
 \right|_{\theta = 0},
\label{eq:action}
\end{eqnarray}
where $\mathcal{L}^{(2)}$ is 
a real superconformal projective 
superfield with weight 2.
The line integral is evaluated over a closed contour $\gamma$ in $\mathbb{C}P^1$. 
Since the action \eqref{eq:action} is independent of $u$, we can choose a frame where $u_i = (1,0)$.
We take the contour $\gamma$ in \eqref{eq:action} such that 
it does not pass through the north pole $v^i=(0,1)$. 
We then introduce a complex inhomogeneous coordinate 
$\zeta \in \mathbb{C}$ in the upper hemisphere of $\mathbb{C}P^1$,
\begin{eqnarray}
v^i = v^1 (1, \zeta), \quad \zeta \equiv \frac{v^2}{v^1}, \quad i=1,2.
\end{eqnarray}
Then the supercovariant derivative $D^{(2)}_{\alpha}$ is rewritten as 
\begin{eqnarray}
 D^{(2)}_{\alpha} = (v^1)^2 D^{[2]}_{\alpha},  \qquad 
  D^{[2]}_{\alpha} (\zeta) \equiv - \bar{\mathbb{D}}_{\alpha} - 2 \zeta
 D^{12}_{\alpha} + \zeta^2 \mathbb{D}_{\alpha}.
\label{eq:cov_deri}
\end{eqnarray}
By factoring out the $v^1$ dependence in $Q^{(n)} (z,v)$, 
a new superfield $Q^{[n]} (z,v)\propto Q^{(n)} (z,v)$ 
is defined as 
\begin{eqnarray}
D^{[2]}_{\alpha} (\zeta) Q^{[n]} (z, \zeta) = 0, \qquad Q^{[n]} (z,
 \zeta) = \sum_k \zeta^k Q_k (z), 
\label{Pconstraint}
\end{eqnarray}
where $Q_k (z)$ are standard $\mathcal{N} = 3$ superfields subject
to the constraints.
Then the action \eqref{eq:action} reduces to the following form,
\begin{eqnarray}
S = \frac{1}{2\pi i} \oint_{\gamma} \! \frac{d\zeta}{\zeta} 
\int d^3 x d^4 \theta \left. \mathcal{L}^{[2]} (z, \zeta)
\right|_{\theta_{12} = 0},
\label{N2action}
\end{eqnarray}
where we have used \eqref{eq:cov_deri} and the constraint
\eqref{Pconstraint}. 
Here the symbol $|_{\theta_{12} = 0}$ means that the 
superfields in the Lagrangian are projected on the $\mathcal{N} = 2$ superspace.
Performing the $\zeta$ integration, we obtain the
action in the standard $\mathcal{N} = 2$ superspace. 

\subsection{$\mathcal{N} = 4$ projective superspace}
For the $\mathcal{N} = 4$ projective superspace, we introduce a pair of
$\mathbb{C}P^1$ \cite{KuPaTaUn}. 
The complex projective spaces $\mathbb{C}P^1_L \times \mathbb{C}P^1_R$ 
are parametrized by the homogeneous complex coordinates $v_L = (v^i), v_R =
(v^{\bar{k}})$ and $u_L = (u_i), u_R = (u_{\bar{k}})$. 
They satisfy the completeness relation \eqref{completeness} independently.
The $\mathcal{N} = 4$ supercovariant derivatives are defined by 
\begin{eqnarray}
\begin{aligned}
 & D^{(1) \bar{k}}_{\alpha} = v_i D^{i \bar{k}}_{\alpha}, \quad
D^{(-1) \bar{k}}_{\alpha} = \frac{1}{(v_L, u_L)} u_i
D^{i\bar{k}}_{\alpha}, \\
 & D^{(1)i}_{\alpha} = v_{\bar{k}} D^{i \bar{k}}_{\alpha}, \quad 
D^{(-1) i}_{\alpha} = \frac{1}{(v_R, u_R)} u_{\bar{k}} 
D^{i \bar{k}}_{\alpha}.
\end{aligned}
\end{eqnarray}
In the $\mathcal{N} = 4$ case, 
one introduces the left and right 
projective superfields with weight $n$ independently.
They are defined by
\begin{eqnarray}
\begin{aligned}
 & D^{(1) \bar{k}}_{\alpha} Q^{(n)}_L (v_L) = 0, \qquad 
Q^{(n)}_L (c v) = c^n Q^{(n)}_L (c v), \\
 & D^{(1) i}_{\alpha} Q^{(n)}_R (v_R) = 0, \qquad 
Q^{(n)}_R (c v) = c^n Q^{(n)}_R (c v), \qquad c \in 
\mathbb{C}^{*}.
\label{constN4}
\end{aligned}
\end{eqnarray}
Since the left and right parts have almost the same property, we focus
on the left part in the following.
We introduce the complex inhomogeneous coordinate
$\zeta_L$ by  
\begin{eqnarray}
v^i = v^1 (1, \zeta_L), \quad \zeta_L = \frac{v^2}{v^1}.
\end{eqnarray}
Then the supercovariant derivative becomes
\begin{eqnarray}
D^{(1)\bar{k}}_{\alpha} = v^1 D^{[1]\bar{k}}_{\alpha}, \qquad 
D^{[1]\bar{k}}_{\alpha} = D_{\alpha}^{2\bar{k}} - \zeta_L D^{1 \bar{k}}_{\alpha}.
\end{eqnarray}
As 
for the $\mathcal{N} = 3$ case, 
the $v^1$ dependencies of the projective superfields 
can be factored out and one can define a new field 
$Q^{[n]}_L \propto Q^{(n)}_L$ which satisfies the following condition,
\begin{equation}
D^{[1]\bar{k}}_{\alpha} (\zeta) Q^{[n]}_L = 0, \qquad Q^{[n]}_L (z',
 \zeta_L) = \sum_k \zeta_L^k Q_k (z'), 
\label{N4Lconstraint}
\end{equation}
where $Q_k(z')$ are the standard $\mathcal{N} = 4$ superfields
subject to the constraint \eqref{constN4}. 

The manifestly $\mathcal{N} = 4$ superconformal invariant action 
is given by 
\begin{eqnarray}
S = 
\frac{1}{2\pi} \oint_{\gamma_L} \! 
(v_L, dv_L)
\int \! d^3 x \ D^{(-4)}_L \mathcal{L}_L^{(2)} (z', v_L) |_{\theta = 0} 
+
\frac{1}{2\pi} \oint_{\gamma_R} \! 
(v_R, dv_R)
\int \! d^3 x \ D^{(-4)}_R \mathcal{L}_R^{(2)} (z', v_R) |_{\theta = 0},
\nonumber \\
\label{N4action}
\end{eqnarray}
where $\mathcal{L}^{(2)}_L(\mathcal{L}^{(2)}_R)$ is 
a left (right) projective superfield with weight 2.
The integration measures are defined by 
\begin{eqnarray}
\begin{aligned}
 & D^{(-4)}_L = \frac{1}{48} D^{(-2) \bar{k} \bar{l}} D_{\bar{k}
 \bar{l}}^{(-2)}, \quad D^{(-2)}_{\bar{k} \bar{l}} = D^{(-1)
 \alpha}_{\bar{k}} 
D^{(-1)}_{\alpha \bar{l}}, \\
 & D^{(-4)}_R = \frac{1}{48} D^{(-2)ij} D_{ij}^{(-2)}, \quad
 D^{(-2)}_{ij} = D^{(-1)
 \alpha}_i D^{(-1)}_{\alpha j}.
\end{aligned}
\end{eqnarray}
The contour $\gamma_L$ $(\gamma_R)$ is chosen such that the path goes
the outside of the north pole in $\mathbb{C}P^1_L$ $(\mathbb{C}P^1_R)$.
After fixing $u_i = (1,0)$, $u_{\bar{k}} = (1,0)$ in $\mathbb{C}P^1_L$
and $\mathbb{C}P^1_R$, the action \eqref{N4action} 
is rewritten in the $\mathcal{N} = 2$ superspace:
\begin{eqnarray}
S = 
\frac{1}{2\pi i} \oint_{\gamma_L} \! \frac{d \zeta_L}{\zeta_L} \int
 \! d^3 x d^4 \theta \mathcal{L}^{[2]}_L (z', \zeta_L)|_{\theta_{\perp} = 0} 
+ 
\frac{1}{2\pi i} \oint_{\gamma_R} \! \frac{d \zeta_R}{\zeta_R} \int
 \! d^3 x d^4 \theta \mathcal{L}^{[2]}_R (z', \zeta_R)|_{\theta_{\perp} = 0},
\end{eqnarray}
where the symbol $|_{\theta_{\perp} = 0}$ means that the 
superfields in the Lagrangian 
are projected on the $\mathcal{N} = 2$ superspace.

\end{document}